\documentclass[aps,prb,twocolumn,showpacs]{revtex4}
\usepackage[T2A,OT1]{fontenc}
\usepackage[cp1251]{inputenc}         % Кодировка документа
\usepackage[english]{babel} % Многоязычность документа
\usepackage[dvips]{epsfig}            % загрузить пакет epsfig и использовать драйвер dvips
\usepackage{latexsym}
\usepackage{amsfonts}
\usepackage{boxedminipage}
\usepackage{longtable}
\begin{document}
%\bibliographystyle{apsrev}
%\preprint{APS/123-QED}
\selectlanguage{english} \sloppy
\title{Magnetostriction-driven multiferroicity of  MnTe and MnTe/ZnTe epitaxial films}
\author{Helen V.Gomonay}
\email{malyshen@ukrpack.net}
\altaffiliation {Bogolyubov Institute for Theoretical Physics NAS
of Ukraine, Metrologichna str. 14-b, 03143, Kyiv, Ukraine}
\author{Ievgeniia G. Korniienko}\affiliation {National
Technical University of Ukraine ``KPI``, ave Peremogy, 37, 03056
Kyiv, Ukraine}
\date{\today}
\begin{abstract}
Here we demonstrate that MnTe epitaxial films with zinc-blend
structure and MnTe/ZnTe multilayers should show ferroelectric
polarization in antiferromagnetically (AFM) ordered state and thus
belong to multiferroics. Spontaneous ferroelectric polarization
results from the bending of highly ionic Mn-Te-Mn bonds induced by
magnetostrictive shear strain. Orientation of ferroelectric
polarization is coupled with orientation of AFM vector and thus
can be controlled by application of the external magnetic field.
Due to the clamping of electric and magnetic order parameters,
domain structure in MnTe is governed by two mechanisms:
depolarizing field produced by electric dipoles and destressing
field produced by magnetoelastic dipoles. The values of
monodomainization electric and magnetic fields depend upon the
sample shape and diminish with the film thickness. Magnetoelectric
nature of the domains make it possible to visualize the domain
structure by linear and nonlinear optical methods (Kerr effect,
second harmonic generation technique).
\end{abstract}
 \maketitle
\section{Introduction}
Multiferroics are a class of a single phase or composite materials
with coexisting magnetic and ferroelectric ordering. High
susceptibility of multiferroics to both electric and magnetic
fields which enables electrical control of magnetic state and vice
versa makes them interesting from fundamental and applied points
of view.

Some of multiferroics are ferroelectric and ferromagnetic (see,
e.g., Refs. \onlinecite{Kimura:2003, Lunkenheimer:2005,
Fert:2007}) and could be used as a storage media with high
information capacity or as functionalized materials for electronic
devices. Others (mainly rare earth compounds with Mn ions like
RMnO$_3$ or RMn$_2$O$_5$, where  R = Tb, Dy, Ho, Y, etc.) show
ferroelectric and antiferromagnetic ordering \cite{Vitebskii:1997,
Fiebig:2004, baier:2006, kundys:2008, okamoto-2007} and may find
an application as mediators for indirect electric control of
ferromagnetic state of an adjacent layer.\cite{Chu:2007}

Cross-coupling between the magnetic and ferroelectric properties
may be caused by different mechanisms. In type-I multiferroics
(according to classification proposed in the recent review
\cite{Brink:2008}) ferroelectricity and magnetism have different
origin and coexist only in a certain temperature range. Materials
classified as type-II show ferroelectricity only in a magnetically
ordered state. Until now such a combined ferroelectric-magnetic
ordering was observed in crystals which posses at least small
noncompensated magnetization, i.e. in ferro-, ferri- and weak
ferromagnets (see, for example, Refs.\ \onlinecite{Morishita:2001,
Vitebskii:1997, Kornev:2000, Fiebig:1998}, etc.).

Two microscopic mechanisms of the magnetically-induced
ferroelectricity originate from the exchange interactions
\cite{Mostovoy:2008}, namely, i) anisotropic exchange
\cite{sergienko-2006-73, hu:2008} (Dzyaloshinskii-Moria
interaction, DMI) and ii) exchange striction \cite{chapon:2006}
(magnetoelastic coupling to the lattice). In the first case
antisymmetric DMI activated by noncollinear spin ordering breaks
inversion symmetry and induces electric polarization through the
concomitant lattice and electronic distortion. Some typical
examples of the compounds which show DMI are listed in
Table~\ref{data_ferroelectric}.

The second mechanism may play the dominant role in the
antiferromagnets with superexchange interactions between the
magnetic ions mediated by bridging via nonmagnetic anions, like
DyMn$_2$O$_5$ and YMn$_2$O$_5$, see
Table~\ref{data_ferroelectric}. In this case the value of exchange
integral strongly depends upon an angle between anion-cation
bonds. In the crystals with the competing antiferromagnetic (AFM)
interactions even small bond-bending may reduce the exchange
energy, stabilizes long-range AFM ordering and produce nonzero
electric polarization.

In the present paper we predict multiferroicity caused by
magnetoelastic mechanism in the epitaxially grown MnTe/ZnTe films
and heterostructures that belong to the family of II-VI
semiconductors \cite{Furdyna:1988} well suited to many
optoelectronic applications in the infrared and visible range
\cite{Pautrat:1994}. Our hypothesis is based on the following
facts.

\begin{enumerate}
  \item  Epitaxial layers of MnTe, ZnTe, CdTe  and corresponding heterostructures posses zinc-blend (ZB) structure
consistent with piezoelectric activity. In particular, isomorphous
to MnTe nonmagnetic films of ZnTe, CdTe, ZnTe/CdTe(111) can
produce a macroscopic electric polarization when stressed or
strained \cite{Andre:1990}.
  \item Predominant mechanism of the magnetic interactions between
  Mn$^{2+}$ ions is superexchange via Te anions \cite{Larson:1988, Furdyna:1988,
  Krenn:1998} that favours antiferromagnetic ordering and
  noticeably varies with Mn-Te-Mn bond bending.
  \item Magnetoelastic coupling in MnTe is rather strong as can be deduced from
step-wise  variation of lattice parameters at the N\'{e}el
temperature. \cite{Giebultowicz:1993(2),
  Szuszkiewicz:2002}
  \item MnTe is a wide-gap (band gap is 3.2 eV, as reported in Ref.\ \onlinecite{Furdyna:1988})
  semiconductor with vanishingly small concentration of the \emph{mobile} carriers at low
  temperatures. \cite{Giebultowicz:2001} Thus, ferroelectric
  polarization should not be seriously affected by screening by the mobile
  charge. In  MnTe/ZnTe heterostructures such a screening can be
  avoided by a proper choice of the superlattice parameters and/or
  by doping.
\end{enumerate}

We argue that the value of spontaneous ferroelectric polarization
in MnTe can be as large as 60~nC/cm$^2$ and thus is comparable
with polarization of many other antiferromagnets (see
Table~\ref{data_ferroelectric}). Moreover, due to the direct
coupling between magnetic and ferroelectric ordering, one should
expect strong magnetoelectric effects (i.e., induction of
polarization by a magnetic field) in this material. Magnetoelastic
mechanism is responsible also for the nonlinear optical effects
and opens a possibility to visualize antiferromagnetic (and
corresponding ferroelectric) domains by use of second harmonic
generation (SHG) technique (see, e.g, Refs.\
\onlinecite{Fiebig:2002, Fiebig:2005}).

The paper is organized as follows. In Sec.\ref{exper} we describe
crystal and magnetic structure of MnTe/ZnTe films and in
Sec.\ref{intuitive} give some intuitive reasons for appearance of
spontaneous electric polarization below the N\'{e}el point. In
Sec.\ref{calc} we calculate the value of spontaneous polarization
on the basis of phenomenological model. Sec.\ref{domain_structure}
is devoted to the discussion of AFM and ferroelectric domains and
methods of the domain structure control. In Sec.\ref{magnetooptic}
we discuss possible magnetooptical effects that could be helpful
in visualization of the magnetic and ferroelectric structure of
MnTe. In the last Sec. the summary of the results obtained is
given.

\begin{table*}\label{data_ferroelectric}
%  \centering
  \caption{Values of spontaneous electric polarization, $P^{\rm (spon)}$, N{\'e}el temperature, $T_N$,  type of AFM ordering (MO), microscopic mechanism and type (according to classification Ref.\onlinecite{Brink:2008}) of ferroelectricity for some multiferroics (MF).}
 \begin{ruledtabular} \begin{tabular}{ccccccc}
Compound&$P^{\rm (spon)}$ (nC/cm$^2$)&$T_N$ (K)&Type of MO&
Mechanism&Type of MF& Source
\\\hline BiFeO$_3$&12$\cdot 10^4$&673&cycloid&octahedra
rot.&I&Ref.\onlinecite{Chu:2007}
\\ BiSrFeO$_3$&10$\cdot 10^4$&643&noncollinear&DMI&I&Ref.\onlinecite{kundys:2008(2)}
\\ DyMn$_2$O$_5$
&150&39&noncollinear&ME&II&Ref.\onlinecite{Higashiyama:2004,
cruz:2006}
\\
YMn$_2$O$_5$&100&45&collinear&ME&II&Ref.\onlinecite{chapon:2006}
\\ MnTe (ZB)&60&65&collinear&ME&II&this work
\\ TbMnO$_3$
&50&27&spiral&DMI&II&Ref.\onlinecite{yamasaki-2007}
\\ TbMn$_2$O$_5$
&40&37&noncollinear&DMI&II&Ref.\onlinecite{okamoto-2007}
\\ Ni$_3$V$_2$O$_8$
&12.5&6.3&noncollinear&DMI&II&Ref.\onlinecite{lawes-2005-95,
kharel-2007}
\\ MnWO$_4$
&4&12.7&noncollinear&DMI&II&Ref.\onlinecite{kundys:2008}
\\ Ca$_3$CoMnO$_6$
&-&13&collinear, Ising&ME&II&Ref.\onlinecite{wu:2008}
\\
\end{tabular}\end{ruledtabular}
\end{table*}

\section{Structure and magnetic properties of MnTe}\label{exper}
\begin{figure}[htbp]
\mbox{\epsfig{file=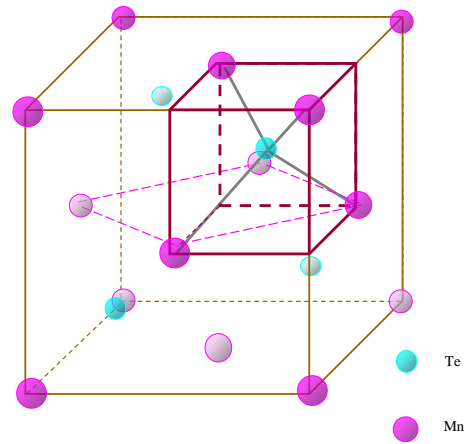,width=0.7\columnwidth}}\caption{(Color
online) Structure of zinc-blend MnTe. Mn - magenta (large)
spheres, Te - blue (small) spheres. Inset (small cube) shows
orientation of Mn-Te bonds. }\label{fig_unit_cell_2}
\end{figure}

\begin{figure}[htbp]
\mbox{\epsfig{file=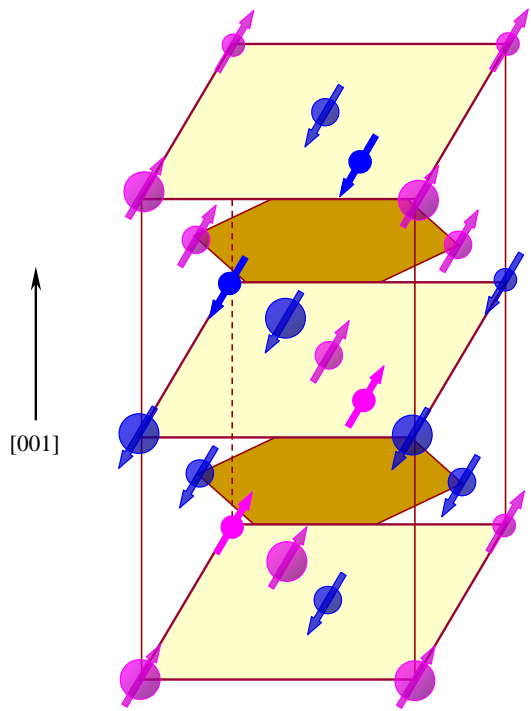,width=0.8\columnwidth}}\caption{(Color
online) Collinear AFM-III structure in the tetragonally distorted
fcc spin lattice. Film growth direction is parallel to [001]
crystallographic axis. Mn$^{2+}$ ions with opposite spin direction
are shown in different colours (opposite
arrows).}\label{fig_magnetic_unit_cell}
\end{figure}

Bulk-like films of paramagnetic MnTe grown by MBE technique have
zinc-blend structure \cite{Giebultowicz:1993(1)} shown in
Fig.~\ref{fig_unit_cell_2} (symmetry group is F$\bar{4}$3m). The
$d-d$ exchange interaction between the
 nearest localized Mn$^{2+}$ spins ($S=5/2$)
 is accomplished via Mn-Te-Mn bonds \cite{Larson:1988} and thus is antiferromagnetic. In the
 strain-free fcc lattice AFM exchange between the nearest neighbors (NN) is
 frustrated.
 Nevertheless, neutron diffraction experiments \cite{Giebultowicz:1993(1), Giebultowicz:1993(2)}
 reveal appearance of the long-range AFM type-III structure below the N\'{e}el temperature
 ($T_N=65$~K). As shown in Fig.~\ref{fig_magnetic_unit_cell}, this
 spin arrangement consists of AFM sheets parallel to (001) plane (represented in the plot by the shaded parallelograms).
Stabilization of long-range ordering is attributed to the
influence of weak NNN coupling \cite{Giebultowicz:1993(2)} and to
misfit-induced strain of the magnetic layer \cite{Rupprecht:2001}
which in the case of MnTe/ZnTe corresponds to elongation
\cite{Giebultowicz:1993(2)} ($c/a=1.06$) in the direction of film
growth.

It should be noted that in the described AFM-III structure the
coupling energy between the spins in the adjacent (001) sheets
sums to zero, so, mutual rotation of spins in different layers
does not change the energy of exchange
interactions\footnote{~Experiments of Hennion et al
\cite{Szuszkiewicz:2002} ruled out the existence of anisotropic
exchange mechanism in this system.}. Stabilization of the
\emph{collinear} AFM-III structure can be due to the pronounced
magnetostriction \cite{Giebultowicz:1993(2)} (0.3\%) that removes
degeneracy between [100] and [010] directions within the film
plane.

It should be also mentioned that the symmetry of AFM state
(corresponding group is generated by the rotation 2$_{[001]}$ and
translation [$\frac{1}{2},\frac{1}{2}$,0] both combined with time
inversion $1^\prime$) allows the existence of macroscopic electric
polarization vector oriented along the [001] axis and forbids
existence of macrosopic magnetization.

\section{Multiferroicity induced by magnetostriction: intuitive considerations}\label{intuitive}
The role of magnetoelastic coupling in the formation of AFM-III
structure can be traced from the following qualitative
considerations. It was already mentioned that the value of
Mn-Te-Mn angle $\phi$ is the key quantity in determining the
exchange coupling constant $J_{d-d}$ between the NN Mn$^{2+}$
ions: upon decreasing the $\phi$ the strength of AFM interaction
increases. For small deflection from an ideal $\phi_0=109.5^\circ$
angle peculiar to fcc lattice, this dependence can be approximated
as follows (from the results of Bruno and Lascaray
\cite{Bruno:1988})
\begin{equation}\label{approximation}
J_{d-d}(\phi)=-3.45+0.135(\phi-\phi_0), \quad \textrm{K}.
\end{equation}
\begin{figure}[htbp] \centering%13772.jpg
\mbox{\epsfig{file=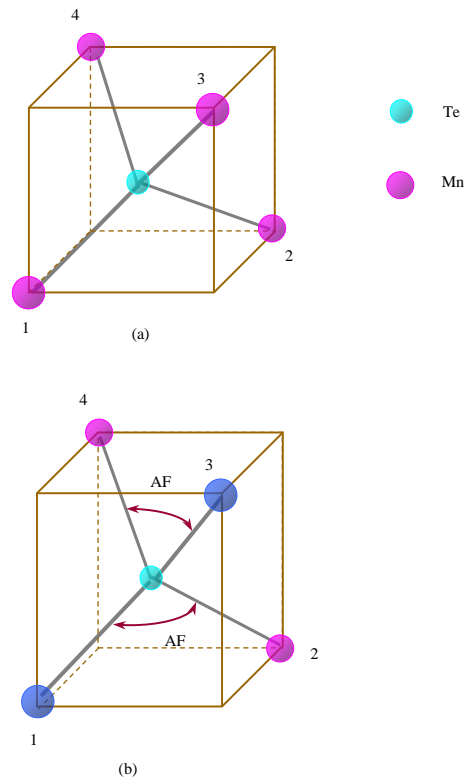,width=0.7\columnwidth}}\caption{(Color
online) Effect of the mismatch-induced stresses. Blue (1, 3) and
magenta (2, 4) spheres correspond to Mn atoms with opposite
directions of magnetization. (a) Nondeformed (cubic) cell, atoms
in 1-4 positions are equivalent and angles between all the Mn-Te
bonds are the same and equal 109.5$^\circ$. (b) Elongation in
[001] direction removes degeneracy between
angles.}\label{fig_unit_cell}
\end{figure}
Now, let us turn to Fig.\ \ref{fig_unit_cell} which illustrates
the effect of tetragonal strain induced by the presence of ZnTe
layers. In a nondeformed cubic lattice all the bonds make an
``ideal'' angle $\phi_0$ and all NN interactions are equivalent.
Tetragonal distortion in [001] direction removes degeneracy
between in-plane (001) and interplane NN exchange. In the case of
elongation the angles for pairs 12, 34 are smaller than those for
the other pairs:
\begin{equation}\label{relations_angle}
  \phi_{12}=\phi_{34}\equiv \phi_{\rm in}<\phi_{13}=\phi_{14}=\phi_{23}=\phi_{24}\equiv \phi_{\rm out}
\end{equation}
Typical values for in-plane, $\phi_{\rm in}$, and interplane,
$\phi_{\rm out}$, angles calculated from geometrical
considerations for some MnTe/ZnTe superlattices are listed in
Table~\ref{data_angles}. In the last two columns of the Table we
give the values of the exchange constants calculated from
Eq.(\ref{approximation}). It can be easily seen that NN exchange
interaction favors  AFM ordering
  for the atoms within (001) plane. At the same time, the bonds
  between the atoms in neighboring planes are still equivalent and
  so are frustrated.

\begin{figure}[htbp] \centering%13772.jpg
\mbox{\epsfig{file=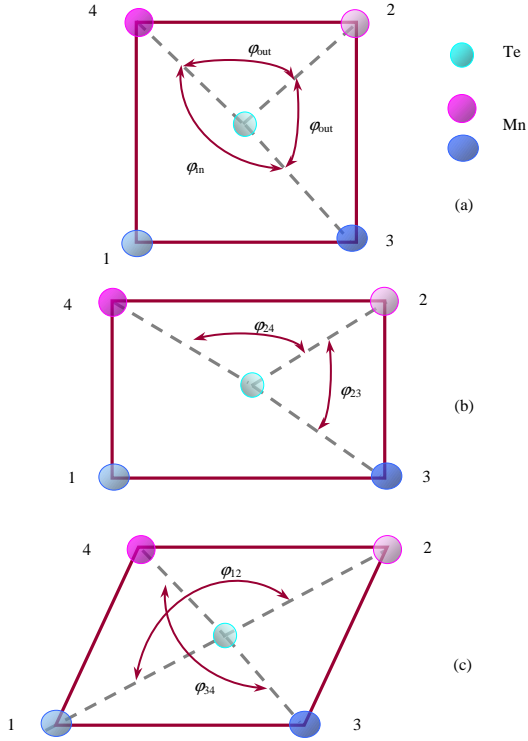,width=0.8\columnwidth}}\caption{(Color
online) Magnetostriction in (001) plane. Pairs of
antiferromagnetically coupled Mn atoms 1, 2 (hollow) and 3,4
(filled)  belong to different atomic planes.(a) Nondeformed state,
angles between Mn-Te-Mn bonds take only two different values:
$\phi_{\rm in}$ (pairs 12 and 34) and $\phi_{\rm out}$ (pairs 13,
14, 23 and 24). (b) Elongation in [100] direction removes
degeneracy between the angles $\phi_{13}=\phi_{24}>\phi_{\rm
out}>\phi_{23}=\phi_{14}$. (c) Shear strain $u_{xy}$ removes
degeneracy between the angles $\phi_{12}$ and
$\phi_{34}$.}\label{fig_deformation_2}
\end{figure}

 Further deformation within (001) plane (see
Fig.\ \ref{fig_deformation_2} a, b) leads to optimization of the
exchange in the interplane FM and AFM bonds by strengthening
exchange interactions with the “right” sign and weakening those
with the “wrong” sign. Namely, elongation in [100] direction
(corresponding strain component
 $u_{xx}-u_{yy}>0$) results in the following difference between the exchange
 constants
\begin{equation}\label{exchange_difference}
  J_{d-d}(\phi_{14})-J_{d-d}(\phi_{13})=\frac{0.27a_{xy}}{\sqrt{a^2_{xy}+c^2}}(u_{xx}-u_{yy}), \quad  \textrm{K},
\end{equation}
where $a_{xy}$ and $c$ are in-plane and interplane lattice
parameters, correspondingly.

Thus, magnetostriction can stabilize an AFM-III structure even in
approximation of NN exchange interactions. On the other hand,
magnetostriction plays an important role in formation of
ferroelectric ordering, as we will show below.

\begin{figure}[htbp] \centering%13772.jpg
\mbox{\epsfig{file=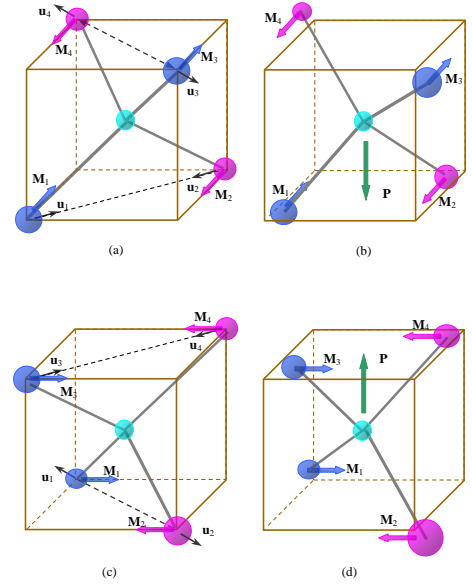,width=0.7\columnwidth}}\caption{(Color
online) Two types of AFM (and ferroelectric) domains. (a), (c)
Mutual orientation of the sublattice magnetization (vectors
$\mathbf{M}_j$, $j=\overline{1,4}$) and shift vector
($\mathbf{u}_j$, $j=\overline{1,4}$) of Mn atom. (b), (d)
Arrangement of atoms after the shear strain. Nonequivalence of 1-2
and 3-4 bonds results in appearance of ferroelectric polarization
$\mathbf{P}$.}\label{fig_deformation}
\end{figure}

In the case of strong spin-orbit coupling some components of
magnetostrictive tensor could depend upon the mutual orientation
of a localized magnetic moment and a direction of anion-cation
bond. This situation in application to MnTe is illustrated in
Figs.~\ref{fig_deformation_2}c, \ref{fig_deformation}. Shear
strain $u_{xy}$ in (001) plane  makes AFM bonds between 1-2 and
3-4 pair inequivalent:
\begin{equation}\label{angle_dif_2}
  \phi_{12}-\phi_{34}=\frac{\sqrt{2}a_{xy}}{c}\frac{3c^2-2a^2_{xy}}{2a^2_{xy}+c^2}u_{xy}\approx 0.6
  u_{xy}.
\end{equation}

The displacement $\Delta \ell$ between the positive (Mn$^{2+}$)
and negative (Te$^{2-}$) ions is thereby equal to
\begin{equation}\label{distance}
  \Delta \ell=\frac{a_{xy}}{9\sqrt{2}}u_{xy}.
\end{equation}
For the totally ionic bond (with the effective charge 2$e$) and
spontaneous strain $u_{xy}=3\cdot 10^{-3}$ such a displacement
induces a local dipole moment $\propto 0.12$~D. Accurate
calculations of ferroelectric polarization induced by spontaneous
magnetostriction will be given in the next Sec.~\ref{calc}.

\begin{table*}
  \caption{Angles between Mn-Te-Mn bonds for in-plane NN ($\phi_{\rm in}$) and interplane NN ($\phi_{\rm out}$) and corresponding values of exchange integrals $J_{d-d}(\phi_{\rm in})$, $J_{d-d}(\phi_{\rm out})$ (K) calculated from geometrical considerations for different superlattices. In-plane,  $a_{xy}$, and interplane, $c$, lattice parameters (\AA) are taken from the experimental paper Ref.\onlinecite{Giebultowicz:1993(2)}.}\label{data_angles}
 \begin{ruledtabular} \begin{tabular}{ccccccc}
 Type&$a_{xy}$&$c$&$\phi_{\rm in }$&$\phi_{\rm out
}$&$J_{d-d}(\phi_{\rm in})$&$J_{d-d}(\phi_{\rm out})$\\\hline
  MnTe10/ZnTe18 &6.130&6.570&105.7&111.4&  -3.20      &
  -3.97\\
MnTe20/ZnTe18&   6.183  & 6.505  & 106.7 &110.9& -3.27& -3.84\\
MnTe130/ZnTe330 & 6.150 &   6.470&106.7& 110.9& -3.28& -3.83\\
 MnTe&   6.346&   6.346&   109.5& 109.5& -3.45&-3.45\\
\end{tabular}
\end{ruledtabular}
\end{table*}

Fig.\ref{fig_deformation} reveals one interesting feature of the
tetragonally distorted  MnTe system: AFM structure can be
implemented in two types of domains that have different
(perpendicular) orientation of local magnetic moments (with
respect to crystal axes) and opposite direction of electric
polarization. AFM domains are clamped with ferroelectric ones.
This means that macroscopic state of a sample can be controlled by
application of either magnetic or electric field (or both).
Behavior of the domain structure in the presence of the external
fields will be discussed in Sec.~\ref{domain_structure}.

\section{Polarization induced by AFM ordering: model and calculations}\label{calc}
Formally, AFM type-III structure can be described by the following
distribution of magnetization vectors ${\mathbf M}({\mathbf r}_j)$
at Mn sites:
\begin{equation}\label{1}
{\mathbf M}({\mathbf r}_j)={\mathbf L}({\mathbf q})\exp(i{\mathbf
{qr}_j}).
\end{equation}
Three equivalent orientations of the structure vector ${\mathbf
q}=(2\pi/a,0,\pi/a)$ (where $a$ is a lattice constant), and two
orientations of ${\mathbf L}$ vector (parallel to [100] and [010]
axes) generate six types of AFM domains in MnTe films. Misfit
strains in the MnTe/ZnTe multilayers reduce degeneracy to two
possible domain types with the same ${\mathbf q}$ and mutually
perpendicular ${\mathbf L}$ vectors in the film
plane.\cite{gomo:2001(4)}

In a single-domain sample with the fixed magnetic order parameter
(or AFM vector) ${\mathbf L}$ the equilibrium values of
magnetostriction tensor, $\textsf{u}$,  and electrical
polarization, ${\mathbf P}$, can be calculated by minimization of
the following expression for the free energy

\begin{widetext}
\begin{equation}\label{2}
  F=\int dV\{[\lambda^ \prime(u_{xx}-u_{yy})+2\lambda_{16}u_{xy}](L_{x}^2-L_{y}^2)\theta(z)+\frac{1}{2}u_{ij}c_{ijkl}u_{kl}\nonumber\\
  + \frac{{\mathbf P}^2}{2\epsilon_0\kappa}+\frac{2e_{14}(z)}{\epsilon_0\kappa}(P_xu_{yz}+P_yu_{zx}+P_zu_{xy})\}
\end{equation}
\end{widetext}
constructed from symmetry considerations. Parameters $\lambda$ are
magnetoelastic constants related to the shear strain, $\textsf{c}$
is a tensor of the elastic modula characteristic to cubic
structure. Electrical properties are described by  low-frequency
dielectric constant $\epsilon_{\rm pm}=\kappa+1$ and piezoelectric
coefficient $e_{14}$, $\epsilon_0$ is the vacuum permittivity.
Periodical alteration of the AFM (MnTe)
 and nonmagnetic (ZnTe) layers in a heterostructure is
described by a form-function $\theta(z)$ which equals 1 in AFM
layer and vanishes in a nonmagnetic spacer.

In the case of ideal interfaces the strain induced by AFM ordering
is homogeneous throughout the AFM and nonmagnetic layers and thus
depends on the ratio of AFM layer thickness $d$ to the period of
superstructure $D$. In particular, macroscopic shear strain
component $u_{xy}$ which is responsible for polarization effect is
given by the following expression
\begin{equation}\label{3}
  u^{\rm (spon)}_{xy}=-\frac{\lambda_{16}M_0^2\rho}{2c_{44}}\frac{d}{D},
\end{equation}
where $\rho=\pm1$ distinguish between two orthogonal orientations
of ${\mathbf L}\|$[100] and [010] in different domains,
$M_0=|{\mathbf{L}}|$ is sublattice magnetization.

Strain-induced contribution\footnote{The analogous contribution
into polarization may also arise from the direct magneto-electric
coupling omitted in the Expr.(\ref{2}).} into spontaneous
polarization calculated from (\ref{2}), (\ref{3}) is
\begin{eqnarray}\label{polarization_spon}
  P^{\rm (spon)}_z=-2e_{14}u^{\rm (spon)}_{xy}=\frac{e_{14}\lambda_{16}M_0^2\rho}{c_{44}}\frac{d}{D}.
\end{eqnarray}
For the multilayered structure the piezoelectric coefficient $e_{14}(z)$ should be
averaged over the AFM/NM bilayer.

It can be easily seen from the Expr.~(\ref{polarization_spon})
that the direction of vector ${\mathbf P}$ correlates with that of
AFM vector ${\mathbf L}$ (factor $\rho$) and takes an opposite
direction in the domains with ${\mathbf L}\|$[100] and [010], in
accordance with intuitive predictions given above (see
Fig.~\ref{fig_deformation}). So, coupling between the
ferroelectric and antiferromagnetic order is accomplished by the
rigid alignment of the antiferromagnetic axis perpendicular to the
polarization direction.

Characteristic value of the spontaneous ferroelectric polarization
in a single-domain  state can be evaluated from
(\ref{polarization_spon}) and available experimental data (see
Table~\ref{data_MnTe})
%(for Cd$_{0.48}$Mn$_{0.52}$Te $e_{14}$=0.097 C/m$^2$, $\epsilon_{\rm pm}$=9.5, $c_{44}$=18.2 GPa
%\Ref.\onlinecite{Maheswaranathan:1985(2)}, $\lambda_{16}$ is
%estimated from spontaneous striction \cite{Giebultowicz:1993(2)})
as $P^{\rm (spon)}= 60$ nC/cm$^2$. For the thick MnTe film $P^{\rm
(spon)}$ can be even higher because Mn atoms cause an enhancement
of electromechanical coupling in zinc-blende structures.
\cite{Maheswaranathan:1985(2)} It is remarkable that among the
materials with comparable ferroelectric polarization (see
Table~\ref{data_ferroelectric}) MnTe shows the highest N{\'e}el
temperature.

The value of the internal electric field $E^{\rm (spon)}\sim$ 80
kV/cm corresponding to the spontaneous polarization of MnTe is, in
turn, close to strain-generated electric field in non-magnetic
III-V heterostructures. \cite{Smith:1997} It seems to be large
enough to be detected due to the change in the photoluminescence
spectra, like it was done in the Ref.~\onlinecite{Andre:1990}.

\section{Switching of polarization by external magnetic/electric field}\label{domain_structure}
Practical applications of multiferroics in information technology
are based on their ability to maintain single ferroelectric domain
state for a long time and change it under application of the
external field. This can be achieved, for example, by using
materials with a high ferroelectric Curie temperature and a robust
large polarization (say, RMnO$_3$, R=Sc, Y, In, Ho–Lu, see Refs.\
\onlinecite{Fiebig:2006, Hur:2008}) that, in turn, gives rise to a
large value of switching field.  In the systems with strong
coupling between polarization and AFM order a single domain state
can be easily fixed by a proper choice of the sample shape, even
for small polarization. Moreover, if AFM and ferroelectric domains
are intimately related and match up spatially (as, e.g., should be
in MnTe), electric polarization of the sample can be switched by
either magnetic or electric field. In this section we analyze
field dependence of macroscopic polarization for different sample
shapes.

Existence of equilibrium domain structure in the ferroics is
usually attributed to the presence of long-range dipole-dipole
interactions between the physically small ordered regions of the
sample. Quantitative description of the phenomena is based on the
account of shape-dependent contribution (stray energy) into free
energy of the sample. In the case of ferroelectric (or
ferromagnetic) materials this contribution can be written as
\begin{equation}\label{ferrolelectric_dipole}
  \Phi_{\rm f-e}=\frac{V}{2}\langle P_j\rangle N_{jk}\langle P_k\rangle,
\end{equation}
where brackets $\langle \ldots \rangle$ mean averaging over the
sample volume $V$ and  $\textsf{N}$ is the second-rank
depolarization tensor the components of which depend upon the
shape of the sample.

Equilibrium domain structure in AFM with nonzero magnetoelastic
coupling can be described in a similar way. Long-range
(destressing) effects originate from the local internal stress
fields $\sigma^{\rm (mag)}_{jk}$ induced by magnetic ordering.
\cite{gomo:2002,gomonay:174439} Corresponding contribution into
free energy of the sample takes a form
\begin{equation}\label{AFM_dipole}
  \Phi_{\rm destr}=\frac{V}{2}\left\{\langle\sigma^{\rm (mag)}_{jp}\rangle\aleph_{jkpt}\langle \sigma^{\rm (mag)}_{kt}\rangle\right\},
\end{equation}
where  $\aleph_{jkpt}$ is the forth-rank destressing tensor which,
like $\textsf{N}$, depends upon the sample shape.

It was already mentioned that the domain structure of MnTe/ZnTe
multilayers is formed by two (out of six) types of the domains
with (\textit{i}) ${\mathbf L}\|[100]$, $\mathbf{P}\|[001]$,
volume fraction $\alpha$, and (\textit{ii}) ${\mathbf L}\|[010]$,
$\mathbf{P}\|[00\overline{1}]$, volume fraction $(1-\alpha)$.
Switching between this two types can be performed by application
of the external electric field $\mathbf{E}$ parallel to [001] axis
or by the magnetic field $\mathbf{H}$ directed along one of the
``easy'' AFM axes ([100] or [010]).

 In the most practical
applications\footnote{~For the ellipsoid-shaped samples
disregarding small contribution from the domain walls and possible
closure domains.)} macroscopic properties (such as polarization or
elongation) of the multidomain sample depend on the single
parameter $0\le\alpha\le1$ that can be calculated from
minimization of shape-dependent ($\Phi_{\rm f-e}+\Phi_{\rm
destr}$) and field-dependent contributions into free energy:
\begin{widetext}
\begin{equation}\label{shape-dependent_energy}
\Phi=V\left\{\left[S_1-E_z P_z^{\rm(spon)}-\frac{\chi
d}{2D}\left(H_y^2-H_x^2\right)\right]\left(\alpha-\frac{1}{2}\right)+\frac{1}{2}S_2\left(\alpha-\frac{1}{2}\right)^2\right\}.
\end{equation}
\end{widetext}
Here spontaneous polarization $P_z^{\rm(spon)}$ within a domain is
defined by the expression (\ref{polarization_spon}), $\chi$ is the
magnetic susceptibility of MnTe and $S_{1,2}$ are shape-dependent
coefficients\footnote{~We use Voight notations for the components
of the symmetric 2-nd and 4-th rank tensors}:
\begin{eqnarray}\label{coeff_S}
  S_1&=&\lambda^\prime M_0^2\sigma^{\rm (mf)}(\aleph_{11}-\aleph_{22}),\nonumber\\
  S_2&=&4N_{3}\left(P_z^{\rm(spon)}\right)^2\\
  &+&M_0^4[(\lambda^\prime)^2(\aleph_{11}+\aleph_{22}-2\aleph_{12})+16\lambda_{16}^2\aleph_{66}],\nonumber
\end{eqnarray}
where $\sigma^{\rm (mf)}\equiv\sigma^{\rm (mf)}_{xx}=\sigma^{\rm
(mf)}_{yy}$ are isotropic stresses induced in the film plane due
to the lattice mismatch.

Analysis of Eq.(\ref{shape-dependent_energy}) shows that a single
domain state of a sample ($\alpha=1$ or 1) can be achieved either
by the proper choice of the shape ($2S_1\ge S_2$) or application
of the external field ($E\ge E_{\rm MD}$ or $H\ge H_{\rm MD}$)
where we've introduced characteristic monodomainization fields as
follows:
\begin{equation}\label{mono_fields}
  E_{MD}=\frac{S_2}{2P_z^{\rm(spon)}}, \quad
  H_{MD}=\sqrt{\frac{S_2D}{\chi d}}.
\end{equation}

In the multidomain state macroscopic ferroelectric polarization
 depends upon the external
fields in a following way
\begin{equation}\label{field_dependence}
\langle
P_z\rangle=P_z^{\rm(spon)}\left(\frac{2S_1}{S_2}-\frac{E_z}{E_{MD}}
-\frac{H_y^2-H_x^2}{H^2_{MD}}\right).
\end{equation}

Let us analyze the properties of the coefficients $S_{1,2}$
assuming that \textit{i}) the sample has the shape of an ellipse
(with semiaxes $a$, $b$ and eccentricity $k=\sqrt{1-b^2/a^2}$)
within the film plane, film thickness is $h$; \textit{ii}) elastic
properties of the material are isotropic ($c_{11}-c_{12}=2c_{44}$)
and are characterized with the shear modulus $c_{44}$ and Poisson
ratio $\nu\equiv c_{12}/(c_{11}+c_{12})$. The destressing and
depolarizing coefficients can be calculated explicitly in two
limiting cases:
\renewcommand{\theenumi}{\roman{enumi}}
\renewcommand{\labelenumi}{\theenumi )}
\begin{enumerate}
  \item ``pillar'' with $h\gg a>b$:
\begin{eqnarray}\label{coeff_pillar}
  S_1&=&\frac{\sigma^{\rm (mf)}d\lambda^\prime M_0^2}{2c_{44}D}\frac{k^2}{(1+\sqrt{1-k^2})^2},\nonumber\\
  S_2&=&\frac{M_0^4d^2}{2c_{44}D^2}\left\{(\lambda^\prime)^2+4\lambda_{16}^2\right.\nonumber\\
  &+&\left.\frac{4\lambda_{16}^2-(\lambda^\prime)^2}{1-\nu}\frac{k^4}{(1+\sqrt{1-k^2})^4}\right\};
\end{eqnarray}

  \item ``thin film'' with $h\ll b<a$:
\begin{eqnarray}\label{coeff_thin_film}
  S_1&=&\frac{h}{b}\cdot\frac{\sigma^{\rm (mf)}\lambda^\prime M_0^2d}{4c_{44}(1-\nu)D}J_2(k),\nonumber\\
  S_2&=&\frac{4}{\epsilon_0\kappa}\left(P_z^{\rm(spon)}\right)^2,
\end{eqnarray}
where the dimensionless shape-factor is given by an integral
\begin{eqnarray}\label{constants}
  J_2(k)&=&\int_0^{\pi/2}\frac{(\sin^2\phi+\cos2\phi/k^2)d\phi}{\sqrt{1-k^2\sin^2\phi}}\nonumber\\
  &\rightarrow&\left\{\begin{array}{cc}
    3\pi k^2/16, & k\rightarrow 0 \\
    1, &  k\rightarrow 1.
  \end{array}\right.
\end{eqnarray}
\end{enumerate}

It is remarkable that coefficient $S_2$ that favors formation of
the domain structure is nonzero in both limiting cases, as can be
seen from (\ref{coeff_S}), (\ref{coeff_pillar}),
(\ref{coeff_thin_film}). This fact can be easily extended to any
geometry. Really, all the depolarizing effects (stray fields) are
related with the flux of the corresponding (ferroelectric,
magnetoelastic, etc) dipole
 moment through the sample
surface. In our case the flux of ferroelectric polarization is
nonzero only through the surface parallel to the film plane, thus
``thin film'' shows strong depolarizing ferroelectric effect while
``pillar'' shows none. In turn, AFM ordering produces stress
dipoles within the film plane, so, corresponding flux is maximal
through the side faces. As a result, strong destressing effect
should be observed for the ``pillar'', not for the ``thin film''.
For the intermediate case of the sample with the developed side
and face surfaces (``ball'' or ``cube'') coefficient $S_2$ is
contributed by both depolarizing mechanisms and so is nonvanishing
for any shape.

Calculations based on the formulas (\ref{coeff_pillar}),
(\ref{coeff_thin_film}) and experimental data from
Table~\ref{data_MnTe} show that in the case of MnTe multiferroic
the destressing effects are much stronger ($S_2\approx
10^5$~J/m$^3$) that the electric depolarization ($S_2\approx
5\cdot10^3$~J/m$^3$), so, monodomainization can be easier achieved
in the ``thin film''.

Another interesting feature of the destressing phenomena is
existence of the effective internal shape-induced field described
by coefficient $S_1$. It arises due to cross-correlation between
isotropic (induced either by substrate or by magnetovolume effect)
and anisotropic internal stresses and has no analog in
ferromagnetic or ferroelectric materials. Coefficient $S_1$, as
seen from (\ref{coeff_pillar}), (\ref{coeff_thin_film}),
(\ref{constants}), depends upon the eccentricity $k$ and vanishes
for the samples isotropic within the film plane ($a=b$).

So, we can deduce that the shape of the sample influences the
domain structure in two ways. First, the field of
monodomainization (\ref{mono_fields}) strongly depends upon the
$h/b$ ratio. Characteristic values of the electric and magnetic
monodomainization fields for two limiting cases ($h\gg b$ and
$h\ll b$) are given in Table~\ref{data_MnTe}. The value of $H_{\rm
MD}$ for the thick sample (``pillar'') is close to the value of
anisotropy field 3.25 T (see Ref.\onlinecite{Swirkowicz:2000}) and
diminishes down to 0.8 T for `thin film''.

Second, single domain state is energetically favorable for the
samples with the overcritical eccentricity (calculated from the
condition $2S_1\ge S_2$). For example, for the thick sample the
critical eccentricity is 0.57 (corresponding aspect ratio
$a/b=1.22$).

\begin{figure}[htbp]
\mbox{\epsfig{file=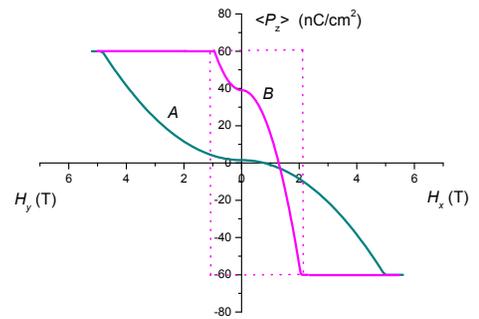,width=0.7\columnwidth}}\caption{(Color
online) Macroscopic ferroelectric polarization as a function of
external magnetic field calculated according to
(\ref{field_dependence}) for ``thin film'' with $h/b=0.1$(curve A,
magenta) and ``pillar'' (curve B, dark cyan). The eccentricity in
both cases is $k=0.1$. Magnetic field $\mathbf {H}$ is switched
between $x$ and $y$ directions. Dotted curve (magenta) shows the
possible field switching between two opposite polarizations in a
single domain sample.}\label{fig_polarization_vs_mag}
\end{figure}

Shape effects are clearly seen in Fig.\
\ref{fig_polarization_vs_mag} which shows field dependence of
macroscopic ferroelectric polarization for the ``thin film'' with
$h/b=0.1$ and for the ``pillar'', both samples having the same
in-plane eccentricity 0.1 (corresponding aspect ratio $a/b=1.005$)
In the first case monodomainization field is smaller and biasing
effect is more pronounced.

\begin{figure}[htbp]
\mbox{\epsfig{file=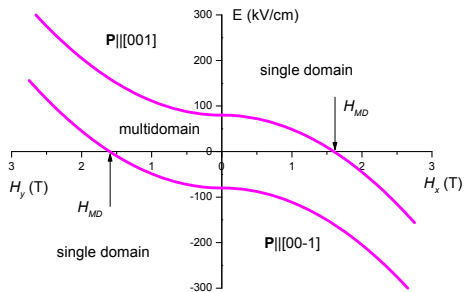,width=0.7\columnwidth}}\caption{(Color
online) Phase diagram in $E-H$ variables for the ``thin film''
($h/b=0.1$) with zero eccentricity. Magnetic field $\mathbf {H}$
is switched between $x$ and $y$
directions.}\label{fig_phase_dia_E_vs_H}
\end{figure}

Monodomainization of the sample can be also achieved by the
combined application of the electric and magnetic field, as
illustrated in Fig.\ \ref{fig_phase_dia_E_vs_H} for the case of
``thin film'' with zero eccentricity. Relation between the
switching fields in this case is given by the formula
\begin{equation}\label{switching_field}
 \left| \frac{E_z}{E_{MD}}
+\frac{H_y^2-H_x^2}{H^2_{MD}}\right| =1.
\end{equation}

\section{Visualization of the domain structure}\label{magnetooptic}
The zinc-blend wide-gap semiconductors are also known to show
nonlinear optical properties studied in second harmonic generation
(SHG) experiments.\cite{Wagner:1998} In the paramagnetic phase
these crystals have the only nonzero SHG tensor component
$d_{14}(-2\omega, \omega, \omega)=d_{25}=d_{36}$ which is rather
large, e.g. for ZnTe it is 119 pm/V at the fundamental wavelength
1047 nm.\cite{Wagner:1998} AFM ordering may bring to being
additional SHG components. To calculate them we analyze the
response to the external electric field with account of the
nonlinear contribution into free energy
\begin{widetext}
\begin{eqnarray}\label{5}
F^{\rm
(nonlin)}&=&-\frac{2d_{14}}{\epsilon_0^2\kappa^3}P_xP_yP_z+\frac{\eta^\prime}{2\epsilon_0\kappa}(P_x^2-P_y^2)(u_{xx}-u_{yy})
+\frac{2\eta_{44}}{\epsilon_0\kappa}(P_xP_yu_{xy}
+P_yP_zu_{yz}+P_zP_xu_{zx})\nonumber\\
&+&\theta(z)[\xi_1(M_x^2+M_y^2)P_z(P_x^2-P_y^2)+\xi_2(M_x^2-M_y^2)P^3_z].
\end{eqnarray}
\end{widetext} Coefficients $\eta^\prime$, $\eta_{44}$ introduced
in (\ref{5}) describe electrostrictive effect and the last term
describes high-order magneto-electric coupling.

Calculations show that in the AFM phase the crystal
 becomes biaxial with anisotropic dielectric
tensor:
\begin{eqnarray}\label{6}
  \epsilon_{xx}&-&\epsilon_{zz}=\epsilon_{zz}-\epsilon_{yy}=\frac{\eta\prime\lambda\prime}{c\prime}M_0^2\rho(\epsilon_{\rm
  pm}-1),\nonumber\\
  \epsilon_{xy}&=&\frac{\eta_{44}\lambda_{16}}{c_{44}}M_0^2\rho(\epsilon_{\rm
  pm}-1),
\end{eqnarray}
in accordance to symmetry prediction for $C_2$ point group of AFM
phase.  The SHG tensor components proportional to AFM vector are
of two types. Like dielectric coefficients, the components
\begin{equation}\label{8}
d_{31}+d_{32}=\frac{4\eta_{44}d_{14}\lambda_{16}}{c_{44}}M_0^2\rho=2u_{xy}^{\rm
(spon)}\eta_{44}d_{14}\rho,
\end{equation}
have opposite signs for the domains with perpendicular orientation
of AFM vectors ($\rho=\pm 1$). The other coefficients are
insensitive to the domain structure:
\begin{eqnarray}\label{7}
  d_{24}&=&-d_{15}=\frac{1}{2}(d_{32}-d_{31})=M_0^2\xi_1\epsilon_0^2(\epsilon_{\rm
  pm}(\omega)-1),\nonumber\\
  d_{33}&=&-M_0^2\xi_2\epsilon_0^2(\epsilon_{\rm
  pm}(\omega)-1).
\end{eqnarray}
The symmetry predicted difference between the $d_{14}$, $d_{25}$
and $d_{36}$ components is due to the high-order terms in the
magnetic order parameter and is neglected.

Rotation of the polarization plane which may stem from anisotropy
of dielectric tensor and  effect of SHG  are sensitive to the
direction of AFM vector and thus open a possibility to visualize
the AFM domain pattern.\cite{Fiebig:2002}

\section{Summary and conclusions}
In the present paper we demonstrate for the first time that MnTe
epitaxial films and MnTe/ZnTe multilayers grown in a convenient
(001)-direction should become electrically polarized below the
N\'{e}el temperature. Spontaneous ferroelectric polarization
results from the bending of highly ionic Mn-Te-Mn bonds induced by
magnetostrictive shear strain.

In contrast to many other multiferroics, MnTe has no macroscopic
magnetic moment and the direction of ferroelectric polarization is
coupled with the orientation of AFM order parameter. Thus, the
domain structure is formed by the combining action of the electric
and magnetoelastic dipole-dipole interactions. Corresponding
depolarizing long-range contribution into free energy of the
sample is nonzero for any sample shape. In the case of thick film
depolarizing effects are governed mainly by magnetoelastic
mechanism and corresponding monodomainization field is of the same
order as characteristic anisotropy field. In the case of thin film
depolarization is due to the presence of the electric dipole
interactions and corresponding monodomainization field is much
smaller.

Due to the clamping between the electrical and AFM domains
switching between the opposite direction of  ferroelectric
polarization can be induced by application of the magnetic field
parallel to ``easy'' AFM axis. Vice versa, one can switch between
different ({100}-oriented and [010]-oriented AFM domains) by
application of the electric field. This effect seems to be useful
for controlling the state of the adjacent ferromagnetic layer
coupled to the multiferroic through exchange interactions like it
was proposed in Ref.\onlinecite{Chu:2007}.

AFM ordering may also induce additional (anisotropy) components of
dielectric and SHG tensors. The sign of the effect is sensitive to
orientation of AFM vectors and thus opens a possibility to
distinguish different domain types. The spontaneous ferroelectric
effect is peculiar to the collinear AFM- III ordering and is not
allowed (from symmetry point of view) in the canted AFM-III
structure (e.g., such an effect is impossible in MnTe/CdTe
multilayers).
\begin{acknowledgements} H.G. would like to acknowledge Prof. P.\ Bruno for fruitful
discussions. This work is supported partly by State Foundation for
Fundamental Researches of Ukraine
%(project N25.2/043)
and partly by Ministry of Science and Education of
Ukraine
%(State registration N 0108U000716).
\end{acknowledgements}
%\bibliographystyle{apsrev}
%\bibliography{MnTe_literatura}

\begin{table}\label{data_MnTe}
  \caption{Values of constants for MnTe (ZB) used in calculations and calculated (see text for notations).}
\begin{ruledtabular}
  \begin{tabular}{cccc}
Constant&Value&Source/formula&Rem. \\\hline
$T_N$&65~K&Ref.\onlinecite{Giebultowicz:1993(2)}&\\
$c/a$&1.06&Ref.\onlinecite{Giebultowicz:1993(2)}&\\
$u_{xx}-u_{yy}$&0,003&Ref.\onlinecite{Giebultowicz:1993(2)}&\\
$e_{14}$&0.097~C/m$^2$&Ref.\onlinecite{Maheswaranathan:1985(2)}&Cd$_{81}$Mn$_{19}$Te\\
$\epsilon_{\rm
pm}$&9,5&Ref.\onlinecite{Maheswaranathan:1985(2)}&\\
$c_{11}$&53.6~GPa&Ref.\onlinecite{Maheswaranathan:1985(2)}&Cd$_{81}$Mn$_{19}$Te\\
$c_{12}$&37~GPa&Ref.\onlinecite{Maheswaranathan:1985(2)}&Cd$_{81}$Mn$_{19}$Te\\
$c_{44}$&18.2~GPa&Ref.\onlinecite{Maheswaranathan:1985(2)}&Cd$_{81}$Mn$_{19}$Te\\
$M_0$&615~Gs&Ref.\onlinecite{Larson:1988}&\\ $\chi$&1.8$\cdot
10^{-4}$ cgs &Ref.\onlinecite{Lewicki:1988}&$T=0$\\
 $\nu$&0.41&$c_{12}/(c_{11}+c_{12})$&\\
 $u_{\rm mf}$&0.03&$0.5(c/a-1)$&\\
$\sigma_{\rm mf}$&2.7~GPa&$(c_{11}+c_{12})u_{\rm mf}$&\\
$\lambda^\prime M_0^2$&0.06~GPa&$c_{44}(u_{xx}-u_{yy})$&\\
$P_z^{\rm (spon)}$&60~nC/cm$^2$&(\ref{polarization_spon})&\\
$E^{\rm (spon)}$&80~kV/cm&$P_z^{\rm (spon)}/(\epsilon_0\kappa)$&\\
$E_{\rm MD}$&750~kV/cm&(\ref{mono_fields})&``pillar''\\ $E_{\rm
MD}$&40~kV/cm&(\ref{mono_fields})&``thin film''\\ $H_{\rm
MD}$&3.5~T&(\ref{mono_fields})&``pillar''\\ $H_{\rm
MD}$&0.8~T&(\ref{mono_fields})&``thin film''\\
\end{tabular}
\end{ruledtabular}
\end{table}
\end{document}